\documentclass[12pt,fleqn]{article}
\usepackage{graphicx}

\usepackage{latexsym}

\usepackage{amsmath}
\usepackage{amsthm}
\usepackage{amssymb}
\usepackage{amsfonts}

\begin{document}

\newcommand{\rf}[1]{(\ref{#1})}
\newcommand{\rff}[2]{(\ref{#1}\ref{#2})}

\newcommand{\ba}{\begin{array}}
\newcommand{\ea}{\end{array}}

\newcommand{\be}{\begin{equation}}
\newcommand{\ee}{\end{equation}}

\newcommand{\const}{{\rm const}}
\newcommand{\ep}{\varepsilon}
\newcommand{\Cl}{{\cal C}}
\newcommand{\rr}{\vec r}
\newcommand{\ph}{\varphi}

\newcommand{\e}{{\bf e}}

\newcommand{\m}{\left( \ba{r}}
\newcommand{\ema}{\ea \right)}
\newcommand{\mm}{\left( \ba{cc}}
\newcommand{\miv}{\left( \ba{cccc}}

\newcommand{\scal}[2]{\mbox{$\langle #1 \! \mid #2 \rangle $}}
\newcommand{\ods}{\par \vspace{0.5cm} \par}
\newcommand{\dis}{\displaystyle }
\newcommand{\mc}{\multicolumn}

\newtheorem{prop}{Proposition}
\newtheorem{Th}{Theorem}
\newtheorem{lem}{Lemma}
\newtheorem{rem}{Remark}
\newtheorem{cor}{Corollary}
\newtheorem{Def}{Definition}
\newtheorem{open}{Open problem}
\newtheorem{ex}{Example}
\newtheorem{exer}{Exercise}

\title{\bf Discrete gradient algorithms of high order for one-dimensional systems}

\author{
 {\bf Jan L.\ Cie\'sli\'nski}\thanks{\footnotesize
 e-mail: \tt janek\,@\,alpha.uwb.edu.pl}
\\ {\footnotesize Uniwersytet w Bia{\l}ymstoku,
Wydzia{\l} Fizyki, ul.\ Lipowa 41, 15-424
Bia{\l}ystok, Poland}
\\ {\bf Bogus{\l}aw Ratkiewicz}\thanks{\footnotesize
e-mail: \tt bograt\,@\,poczta.onet.pl} 
\\ {\footnotesize  I Liceum Og\'olnokszta{\l }c\c ace,
ul.\ \'Sr\'odmie\'scie 31, 16-300 August\'ow, Poland           }
}

\date{}

\maketitle

\begin{abstract}
We show how to increase the order of one-dimensional discrete gradient numerical integrator without losing its advantages, such as exceptional stability, exact conservation of the energy integral and exact preservation of the trajectories in the phase space. The accuracy of our integrators is higher by several orders of magnitude as compared with the standard discrete gradient scheme (modified midpoint rule) and, what is more, our schemes have very high accuracy even for large time steps.  
\end{abstract}

\ods




\pagebreak

\section{Introduction}

In this paper we introduce and develop discrete gradient schemes of high order. 
Discrete gradient schemes are useful tools for numerical integration of many-body dynamical systems \cite{LaG,IA,STW,Gon,HLW}. 
They preserve exactly (up to round-off errors) both the total energy and angular momentum. More recently discrete gradient methods have been extended and developed in the context of geometric numerical integration \cite{MQ}. Quispel and his coworkers constructed numerical integrators preserving all integrals of motion of a given system of ordinary differential equations \cite{QC,QT,MQR1,MQR2}. Similar ideas were applied in molecular dynamics simulations of spin liquids \cite{OMF1}. 

In general, geometric numerical integrators are very good in preserving qualitative features of simulated differential equations but it is not easy to enhance their accuracy (in the same time preserving their geometric properties). Symplectic algorithms can be improved using appropriate splitting methods \cite{Su,FR,Yo,Bl,MQ-split,OMF2}. Our research is concentrated on improving  the efficiency of the discrete gradient method (which is not symplectic) without loosing its outstanding qualitative advantages. Results reported in earlier papers are very promising \cite{CR-long,CR-PRE}. 

In this paper we present a further essential improvement of our approach  constructing discrete gradient schemes of any prescribed order $N$ for 
one-dimensional Hamiltonian systems of the form:
\be  \label{Newton}
\dot p =  - V'(x) \ , \quad \dot x = p \ , 
\ee
where $V(x)$ is a potential, and the dot and the prime denote differentiation with respect to $t$ and $x$, respectively. In this case the  discrete gradient method reduce to the so called modified midpoint rule: 
\be \ba{l} \label{dis-grad} \dis
\frac{x_{n+1} - x_n}{h} = \frac{1}{2} \left( p_{n+1} + p_n \right) \ . \\[3ex] \dis
\frac{p_{n+1} - p_n}{h} =  - \frac{ V (x_{n+1}) - 
V (x_n) }{x_{n+1} - x_n} \ ,
\ea \ee
where $h$ is the time step.

\section{Discrete gradient schemes of $N$th order}
\label{sec-grad-N}

We consider the following family of nonstandard numerical schemes (parameterized by  a single function $\delta$):  
\be \ba{l} \label{graddel} \dis
\frac{x_{n+1} - x_n}{\delta} = \frac{1}{2} \left( p_{n+1} + p_n \right) \ . \\[3ex] \dis
\frac{p_{n+1} - p_n}{\delta} =  - \frac{ V (x_{n+1}) - 
V (x_n) }{x_{n+1} - x_n} \ ,
\ea \ee
where $\delta$ can depend on any variables and parameters, including $h, x_n, p_n$, $x_{n+1}, p_{n+1}$. 
One can easily prove that the total energy is preserved, i.e., 
\be  \label{energy}
 \frac{1}{2} p_n^2 + V (x_n) = E = \const \ , 
\ee
for any choice of the function $\delta$. This is an essential  generalization of the  well known case $\delta = h$.   

In our recent papers \cite{CR-long,CR-PRE,Ci-oscyl} 
we consider $\delta$ of the form  
\be \label{deltan} 
\delta =
\frac{2}{ \omega } \tan\frac{ h \omega  }{2} \ , \qquad \omega = \sqrt{ V'' (\bar x) } \ , 
\ee
where $\bar x$ may depend on $x_n, x_{n+1}$ but usually does not depend on $h$. Taking $\bar x = x_0$, where $V' (x_0) = 0$, we get 
the modified discrete gradient scheme (MOD-GR) \cite{CR-long}. Then,  $\bar x = x_n$ and $\bar x = \frac{1}{2} (x_n + x_{n+1})$ yield locally exact discrete gradient scheme (GR-LEX) and its symmetric modification (GR-SLEX),  see \cite{CR-PRE,Ci-oscyl}. These three numerical methods are of second, third, and fourth order, respectively \cite{CR-PRE}.

In the present paper we will show that the family of numerical integrators of the form \rf{graddel} contains numerical schemes of any order. 
 The explicit formulae will be presented up to the order 11. 

The system \rf{graddel} (where $x_n \equiv x$ and $p_n \equiv p$ are given and $\delta_n \equiv \delta$ is a small parameter) implicitly defines $x_{n+1}$ and $p_{n+1}$.  Therefore, using implicit differentiation, we can write down the corresponding Taylor series:  
\be  \ba{l}  \label{Tay-1}
x_{n+1} = x + p \delta   - \frac{1}{2} V' \delta^2 - \frac{1}{4} p V'' \delta^3 + \frac{1}{24} \left( 3 V' V'' - 2 p^2 V''' \right) \delta^4 + O (\delta^5) \ , \\[2ex]
p_{n+1} = p - V' \delta - \frac{1}{2} p V'' \delta^2 + \frac{1}{12} \left( 3 V' V'' - 2 V''' p^2 \right) \delta^3 \\[2ex] \qquad \  - \frac{1}{24} \left(  4 p V' V'''  + 3 p (V'')^2 - p^3 V^{(4)} \right) \delta^4 + O (\delta^5) \ . 
\ea \ee
Now, we assume that $x_{n+1}$ and $p_{n+1}$ coincide with the exact solution up to the order $N$, i.e., their expansion in Taylor series have at least $N$ first terms identical with the Taylor series \rf{Tay-bc}.  Then, we compute the first $N$ terms of the Taylor series of $\delta$ using the first equation of \rf{graddel}, i.e.,
\be
  \delta = \frac{2 (x_{n+1} - x_n)}{p_{n+1} + p_n} \ .
\ee
The resulting polynomial of $N$th order is denoted by $\delta_N$ and its coefficients are denoted by $a_k$, i.e.,
\be  \label{deltaN}
\delta_N = \delta_N (x,p,h) = \sum_{k=1}^N   a_{k} (x,p) h^k  = h + \sum_{k=3}^N   a_{k} (x,p) h^k 
\ee
where $a_1 = h$, $a_2 = 0$, and $a_k$ (for $k \geq 3$) are polynomials with respect to $p$ with coefficients depending on $x$ through derivatives of $V$. 
Denoting by a subscript the differentiation with respect to $x$ (and using abbreviations like $V_{4x} \equiv V_{xxxx}$) we present some number of coefficients $a_k$ in an explicit form: 
\be
a_3 = \frac{1}{12} V_{xx} \ , \qquad a_4 = \frac{1}{24} p V_{xxx} \ , 
\ee
\be
a_5 = \frac{1}{240} \left( 2 V_{xx}^2 - 4 V_x V_{xxx} + 3 p^2 V_{4x}  \right) \ , 
\ee
\be
a_6 = \frac{1}{1440}  \left( ( 5 V_{xx} V_{xxx} - 15 V_x V_{4x}) p  + 4  V_{5x} p^3 \right)  \ , 
\ee
\be
a_7 = \frac{1}{20160} \left(  a_{70} + a_{72} p^2 + a_{74} p^4 \right) \ , 
\ee
\be  \ba{l}  \dis
a_8 = \frac{1}{40320}  \left( a_{81} p + a_{83} p^3 + a_{85} p^5    
  \right) \ , \\[2ex]\dis
a_9 = \frac{1}{725760}  \left(  a_{90} + a_{92} p^2 + a_{94} p^4 + a_{96} p^6 \right)  \ , 
 \\[2ex]\dis
a_{10} = \frac{1}{7257600}  \left(  a_{101} p + a_{103} p^3 + a_{105} p^5 + a_{107} p^7 \right)  \ , 
\\[2ex]\dis
a_{11} = \frac{1}{159667200}  \left(  a_{110} + a_{112} p^2 + a_{114} p^4 + a_{116} p^6 + a_{118} p^8 \right)  \ , 
\ea \ee
\ods
\noindent  where the coefficents $a_{jk}$, $a_{jkm}$ depend on $x$ through derivatives of $V$, namely: 
\be \ba{l}
a_{70} = 17 V_{xx}^3 + 45 V_x^2 V_{4x} - 44 V_x V_{xx} V_{xxx}  \ , 
\\[2ex] 
a_{72} =  20 V_{xxx}^2 - 12 V_{xx} V_{4x} - 72 V_x V_{5x}  \ , 
\\[2ex] 
a_{74} = 10  V_{6x}  \ , 
\ea \ee
\be  \ba{l}
 a_{81} = 21 V_{xx}^2 V_{xxx} - 42 V_x V_{xxx}^2 + 63 V_x^2  V_{5x} \ , \\[2ex]
a_{83} = 14 V_{xxx} V_{4x} - 21 V_{xx} V_{5x} - 35 V_x V_{6x} \ , \\[2ex]
a_{85} = 3 V_{7x}  \ , 
\ea \ee
\be \ba{l} 
\begin{split}
a_{90} & = 62 V_{xx}^4 - 228 V_x V_{xx}^2 V_{xxx} + 168 ( V_x^2 V_{xxx}^2 - V_x^3 V_{5x} ) \\[1ex] & + 90 V_x^2 V_{xx} V_{4x} \ , 
\end{split} \\[4ex] 
\begin{split}
a_{92} & = 75 V_{xx} V_{xxx}^2 + 81 V_{xx}^2 V_{4x} - 462 V_x V_{xxx} V_{4x} \\[1ex] & + 360 V_x V_{xx} V_{5x} + 420 V_x^2 V_{6x} \ , 
\end{split}
\\[4ex]
a_{94} = 42 V_{4x}^2 - 120 ( V_{xx} V_{6x} +  V_x V_{7x} ) \ , \\[2ex]
a_{96} = 7 V_{8x}  \ , 
\ea \ee
\be \ba{l} 
\begin{split} a_{101} & = 
460 V_{xx}^3 V_{xxx} - 1170 V_x V_{xx} V_{xxx}^2 - 630 V_x V_{xx}^2 V_{4x} \\[1ex] & + 2385 V_x^2 V_{xxx} V_{4x} - 945 V_x^2 V_{xx} V_{5x} - 1260 V_x^3 V_{6x}  \ , 
\end{split}  \\[4ex] 
\begin{split} a_{103} & = 
150 V_{xxx}^3 + 15 V_{xx} V_{xxx} V_{4x} - 945 V_x V_{4x}^2 - 456 V_x V_{xxx} V_{5x}   
\\[1ex] & 
+ 483 V_{xx}^2 V_{5x}  + 1785 V_x V_{xx} V_{6x} + 1080 V_x^2 V_{7x}  \ , 
\end{split} \\[4ex] 
a_{105} = 
126 V_{4x} V_{5x} - 114 V_{3x}V_{6x} - 261 V_{xx} V_{7x} - 189 V_x V_{8x} \ , 
\\[2ex]
a_{107} = 8 V_{9x}  \ , 
\ea \ee
\be \ba{l} 
\begin{split} a_{110} & = 
1382 V_{xx}^5  - 6448 V_x V_{xx}^3 V_{3x} + 4140 V_x^2 V_{xx}^2 V_{4x} 
\\[1ex] & + 840 V_x^3 V_{xx} V_{5x} + 8280 V_x^3 V_{3x} V_{4x} + 7368 V_x^2 V_{xx} V_{3x}^2 \\[1ex] & + 3150 V_x^4 V_{6x}   \ , 
\end{split}  \\[6ex] 
\begin{split} a_{112} & = 
3240 V_{xx}^2 V_{3x}^2 - 6480 V_x V_{3x}^3 + 696 V_{xx}^3 V_{4x} 
\\[1ex] & + 4872 V_x V_{xx} V_{3x} V_{4x} + 15660 V_x^2 V_{4x}^2 - 9144 V_x V_{xx}^2 V_{5x} 
 \\[1ex] & + 11988 V_x^2 V_{3x} V_{5x} - 21000 V_x^2 V_{xx} V_{6x} - 10800 V_x^3 V_{7x}   \ , 
\end{split}  \\[6ex] 
\begin{split} a_{114} & = 
1710  V_{3x}^2 V_{4x} - 1803 V_{xx} V_{4x}^2  - 8676 V_x V_{4x} V_{5x}  
\\[1ex] & 
+ 1260 V_{xx} V_{3x} V_{5x}  + 4770   V_{xx}^2 V_{6x} + 3060 V_x V_{3x} V_{6x}   \\[1ex] & 
+ 11400 V_x V_{xx} V_{7x} + 4725 V_x^2 V_{8x}  \ , 
\end{split} \\[6ex] 
\begin{split}
 a_{116} & = 
336 V_{5x}^2 - 780 V_{3x} V_{7x} - 980 V_{xx} V_{8x} - 560 V_x V_{9x}  \\[1ex] &  + 120 V_{4x} V_{6x}  \ , 
\end{split} \\[4ex] 
\begin{split}
 a_{118} & = 
18 V_{10x}  \ .
\end{split} 
\ea \ee

The numerical scheme \rf{graddel}, where $\delta = \delta_N$ is defined by \rf{deltaN}, will be denoted by GR-$N$. The  method GR-$N$ is of (at least) $N$th order. We point out that $a_1 = 1$, $a_2 = 0$. It means that the methods GR-1 and GR-2 coincide with the discrete gradient method (GR) given by \rf{dis-grad} (in particular, GR-1 is of 2nd order). Actually, if the potential $V$ is linear in $x$, then any method GR-N is exact (i.e., its order becomes infinite).

\section{Explicit Taylor schemes of $N$th order}
\label{sec-N-explicit} 

In this section we derive explicit numerical schemes of any order, using standard Taylor expansions. 
We expand $x (t + h)$ and $p (t + h)$ in Taylor series:
\be
   x (t+h) = \sum_{k=0}^\infty \frac{h^k}{k!} \frac{d^k x (t)}{d t^k} \ , \qquad p (t+h) = \sum_{k=0}^\infty \frac{h^k}{k!} \frac{d^k p (t)}{d t^k} \ , 
\ee
where all derivatives can be replacd by functions of $x, p$ using \rf{Newton} and its differential consequences (e.g., $\ddot p = - V''(x) \dot x = - V''(x) p$). Thus  we get 
\be \ba{l}  \label{Tay}
x (t + h) = x + p h - \frac{1}{2} V' h^2 - \frac{1}{6} p V'' h^3 + \frac{1}{24}  \left( V' V'' - V''' p^2 \right) h^4 + O (h^5) \\[2ex]
p (t + h) = p - V' h - \frac{1}{2} p V'' h^2 + \frac{1}{6} \left( V' V'' - V''' p^2 \right) h^3 \\[2ex] \qquad \ + \frac{1}{24} \left( 3 p V' V''' + p (V'')^2 - p^3 V^{(4)} \right) h^4 + O (h^5) \ .
\ea \ee
Therefore, the Taylor expansion can be represented in the form
\be \label{Tay-bc}
  x (t + h) = \sum_{k=0}^\infty  \frac{h^k}{k!} \ b_k (x, p)  \ , \qquad
  p (t + h) = \sum_{k=0}^\infty  \frac{h^k}{k!} \ c_k (x, p) \ ,
\ee
where $b_k = \frac{d^k}{d t^k} x$, $c_k = \frac{d^k}{d t^k} p$ and we compute these derivative using \rf{Newton}. For instance,  $b_0 = x$, $b_1 = \dot x = p$ and $b_2 = \ddot x = \dot p = - V' (x)$. In general, 
\be \label{brec} 
   b_{k+1} = \frac{d}{d t} b_k = \frac{ \partial b_k }{\partial x} \dot x + \frac{ \partial b_k }{\partial p} \dot p =  p \frac{ \partial b_k }{\partial x} - V' (x) \frac{ \partial b_k }{\partial p} \ .
\ee
Then, $p = \dot x$ implies 
\be  \label{ck}
   c_k = \frac{d}{d t} b_k = b_{k+1} \ .
\ee
The coefficients $b_k$ ($k=1,2,\ldots,11$), computed recursively from \rf{brec}, read 
\be \ba{l}  \label{b1-7}
b_0 = x \ , \quad b_1 = p \ , \quad b_2 = - V_x \ , \quad b_3 = - p V_{xx} \ , \\[2ex]
b_4 = V_x V_{xx} - p^2 V_{xxx} \ , \\[2ex] 
b_5 = p ( V_{xx}^2 + 3  V_x V_{xxx} ) - p^3 V_{4x}  \ , \\[2ex]
b_6 = - 3 V_x^2 V_{xxx} - V_x V_{xx}^2 + p^2 ( 5  V_{xx} V_{xxx} + 6  V_x V_{4x} )  - p^4 V_{5x} \ , \\[2ex]
\begin{split}
b_7  & = - p (V_{xx}^3 + 18 V_x V_{xx} V_{xxx} + 15 V_x^2 V_{4x} ) 
\\[1ex]  & + p^3 ( 5 V_{xxx}^2 + 11 V_{xx} V_{4x} + 10 V_x V_{5x} ) - p^5 V_{6x}  \ , 
\end{split}
\ea \ee
\be  \label{b8}
\begin{split}
b_8  & = V_x V_{xx}^3 + 18 V_x^2 V_{xx} V_{xxx} + 15 V_x^3 V_{4x} 
\\[1ex]  & - p^2 ( 21 V_{xx}^2 V_{xxx} + 33 V_x V_{xxx}^2 + 81 V_x V_{xx} V_{4x} + 45 V_x^2 V_{5x} ) 
\\[1ex] & + p^4 (21 V_{3x} V_{4x} + 21 V_{xx} V_{5x} + 15 V_x V_{6x} ) - p^6 V_{7x} \ ,  
\end{split}
\ee
\be  \label{b9}
\begin{split}
b_9  & = p ( V_{xx}^4 + 81 V_x V_{xx}^2 V_{3x} + 84 V_x^2 V_{3x}^2 + 225 V_x^2 V_{xx} V_{4x} + 105 V_x^3 V_{5x} )  
\\[1ex]  & - p^3 ( 75 V_{xx} V_{3x}^2 + 102 V_{xx}^2 V_{4x} + 231 V_x V_{3x} V_{4x} + 255 V_x V_{xx} V_{5x}  
\\[1ex] & +  105  V_x^2 V_{6x} ) + p^5 (21 V_{4x}^2 + 42 V_{3x} V_{5x} + 36  V_{xx} V_{6x} + 21 V_x V_{7x} ) 
\\[1ex] & - p^7 V_{8x} \ , 
\end{split}
\ee
\be  \label{b10}
\begin{split}
b_{10}  & =  - ( V_x V_{xx}^4 + 81 V_x^2 V_{xx}^2 V_{3x} + 84 V_x^3 V_{3x}^2 + 225 V_x^3 V_{xx} V_{4x} 
\\[1ex]  & + 105 V_x^4 V_{5x} ) + p^2 ( 85 V_{xx}^3 V_{3x} + 555 V_x V_{xx}  V_{3x}^2  + 837 V_x V_{xx}^2 V_{4x} 
\\[1ex] & + 1086 V_x^2  V_{3x} V_{4x} + 1305 V_x^2 V_{xx} V_{5x} + 420 V_x^3 V_{6x} )   
\\[1ex] & - p^4 ( 75 V_{3x}^3 + 585 V_{xx} V_{3x} V_{4x} + 336 V_x V_{4x}^2 + 357 V_{xx}^2 V_{5x} 
 \\[1ex] & + 696 V_x V_{3x} V_{5x} + 645 V_x V_{xx} V_{6x} + 210 V_x^2 V_{7x} ) + p^6 ( 84 V_{4x} V_{5x} 
\\[1ex] & + 78 V_{3x} V_{6x} + 57 V_{xx} V_{7x} + 28 V_x V_{8x} ) - p^8 V_{9x} \ ,
\end{split}
\ee
\be \label{b11}
\begin{split}
b_{11}  & =  -  p ( V_{xx}^5 + 336 V_x V_{xx}^3 V_{3x} + 1524 V_x^2 V_{xx} V_{3x}^2 + 2430 V_x^2 V_{xx}^2 V_{4x} 
\\[1ex]  & + 2565 V_x^3 V_{3x} V_{4x} + 3255 V_x^3 V_{xx} V_{5x} + 945 V_x^4 V_{6x} ) 
\\[1ex] & + p^3 ( 810 V_{xx}^2 V_{3x}^2 + 855 V_x V_{3x}^3  + 922 V_{xx}^3 V_{4x} + 2430 V_x^2 V_{4x}^2 
\\[1ex] & + 4875 V_x V_{xx}^2 V_{5x} + 5175 V_x^2 V_{3x} V_{5x} + 5145 V_x^2 V_{xx} V_{6x} 
\\[1ex] & + 7296 V_x V_{xx} V_{3x} V_{4x}  + 1260 V_x^3 V_{7x} )   - p^5 ( 810 V_{3x}^2 V_{4x} 
 \\[1ex] & + 921 V_{xx} V_{4x}^2  + 1995 V_{xx} V_{3x} V_{5x} + 1872 V_x V_{4x} V_{5x} + 1002 V_{xx}^2 V_{6x} 
\\[1ex] & + 1809 V_x V_{3x} V_{6x} + 1407 V_x V_{xx} V_{7x} + 378 V_x^2 V_{8x} )  + p^7 ( 84 V_{5x}^2 
\\[1ex] & + 162 V_{4x} V_{6x} + 135 V_{3x} V_{7x} + 85  V_{xx} V_{8x} + 36 V_x V_{9x} ) - p^9 V_{10x} \ .
\end{split}
\ee
\ods
Thus for any fixed $N$ we obtained the following explicit numerical scheme,  denoted TAY-$N$ (the Taylor scheme of $N$th order),  
\be \label{TayN}
 x_{n+1} = \sum_{k=0}^N \frac{h^k}{k!} \ b_k (x_n,p_n) \ , \quad p_{n+1} = \sum_{k=0}^N \frac{h^k}{k!} \ c_k (x_n, p_n) \ ,
\ee
where $b_k$ and $c_k$ are defined by \rf{brec}, \rf{ck} and, in particular cases, by \rf{b1-7}, \rf{b8}, \rf{b9}, \rf{b10} and \rf{b11}. Explicit integrators Tay-$N$ will be used for comparison with discrete gradient methods of high order. Moreover, they are good candidates for predictors when gradient methods \rf{graddel} are used as correctors.

\section{Numerical experiments}

In our recent papers we compared several discretizations of the simple pendulum equation ($V (x) = - \cos x$) with a special stress on the long-time behaviour, see \cite{CR-long,CR-PRE}. Locally exact discrete gradient schemes (GR-LEX and GR-SLEX) turned out to be the best. In some tests their accuracy was better by several orders of magnitude in comparison to standard methods like leap-frog, implicit midpoint rule or the discrete gradient method (GR).  GR-LEX and GR-SLEX yield rather similar results and in this section we confine ourselves to GR-LEX only. 

\begin{figure*}
\includegraphics[width=\textwidth]{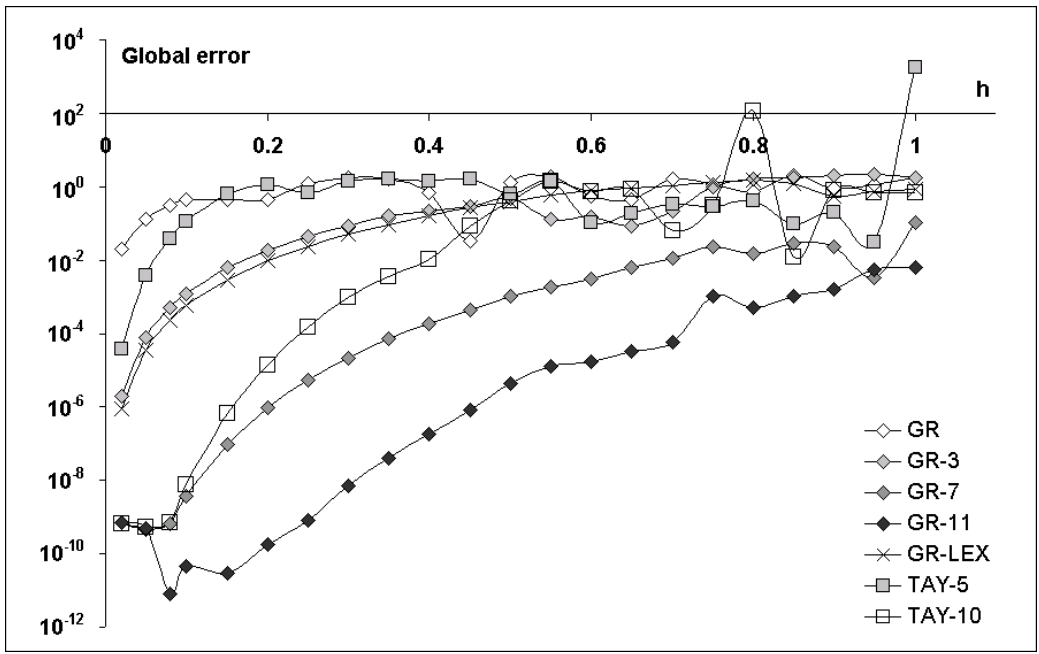}    
\caption{ Global error at $t = 120 T_{th}$ as a function of the time step $h$ for Morse potential, for $p_0 = 0.8$ ($T_{th} = 10.471\ 975\ 51$). }
\label{err-Mor-08}       
\end{figure*}

We are going to compare GR-LEX with algorithms of higher order introduced in the present paper, i.e., GR-$N$ and TAY-$N$. 
The accuracy of these schemes was tested mainly for the simple pendulum, but other potentials yield similar results. We present some data for the Morse potential, $ V(x) = \frac{1}{2} e^{-2x} - e^{-x}$, see Fig.~\ref{err-Mor-08}. In both cases the exact solution is known. For simplicity we always assume the initial position at the stable equilibrium, i.e., $x_0 = 0$.  The details of  numerical computations of the period are explained in \cite{CR-long} and iteration procedures are described and discussed in \cite{CR-PRE} (we apply the fixed point method and the Newton method, and iterate until the acuracy $10^{-16}$ is obtained). We point out that $\delta_N$ given by \rf{deltaN} depends on $x_n, p_n$ and does not depend on $x_{n+1}, p_{n+1}$). It means that $\delta_N$ is evaluated only once at every step.

\subsection{Global error}

\begin{figure*}
  \includegraphics[width=\textwidth]{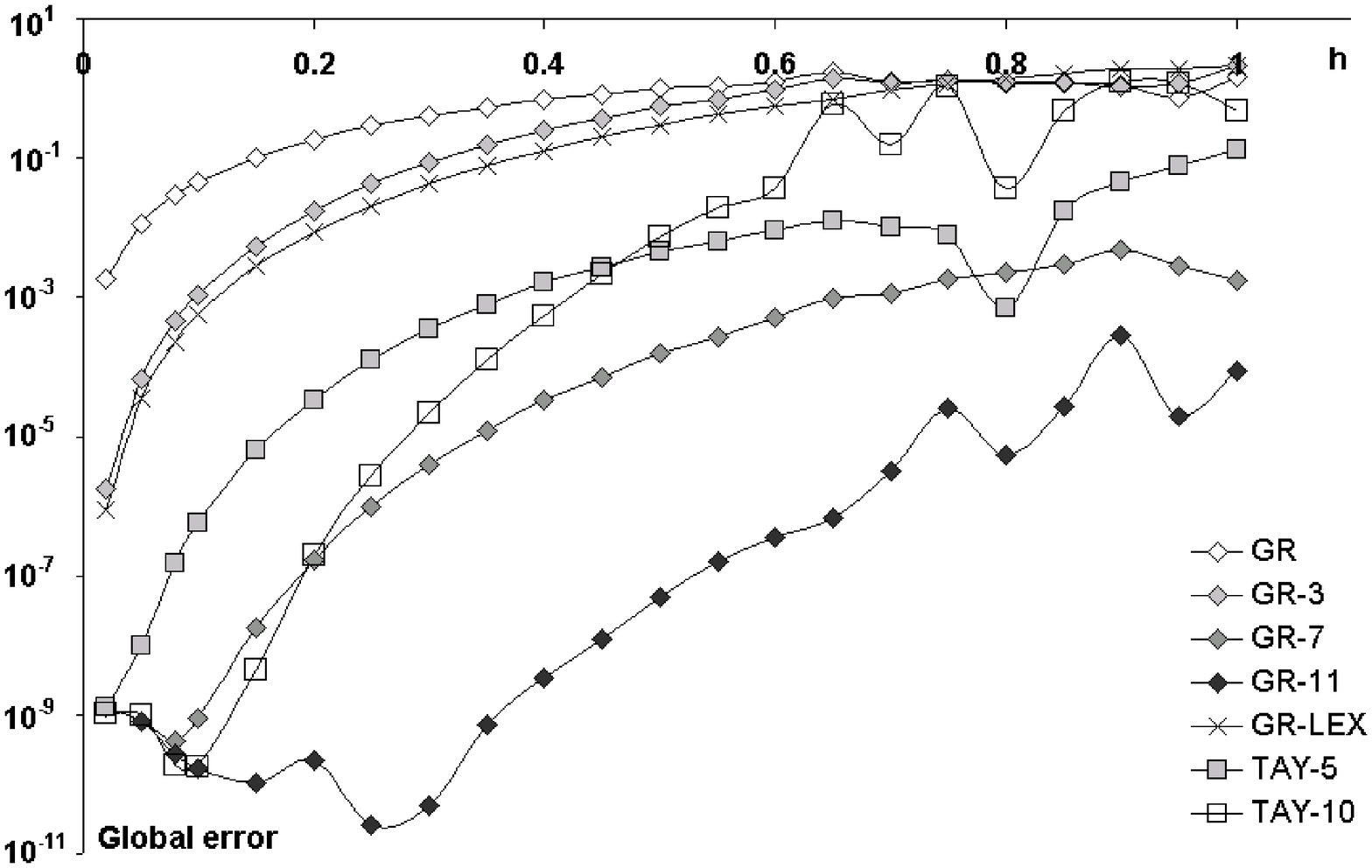}  
\caption{Global error at $t = 120 T_{th}$ as a function of the time step $h$ for the simple pendulum, $p_0 = 1.8$ ($T_{th} = 9.122\ 196\ 55$).  }
\label{err-18}       
\end{figure*}

Fig.~\ref{err-18} and Fig.~\ref{err-Mor-08} show the dependence of the global error of the numerical solutions on the time step (the global error was evaluated at $t = 120 T_{th}$). GR-3 yields almost the same results as GR-LEX. They are better than GR  by several orders of magnitude. GR-$N$ (for $N \geqslant 5$) are more accurate than GR-LEX by several orders of magnitude. 
We point out that the schemes GR-$N$ are very accurate for large time step. Actually, for small time steps (say, $h < 0.1$) the accuracy of GR-7 and GR-11 almost does not depend on $h$ (actually, it even slightly decreases for smaller $h$). TAY-10 becomes less accurate than GR-7 and TAY-5 for larger $h$. 

\begin{figure*}
\includegraphics[width=\textwidth]{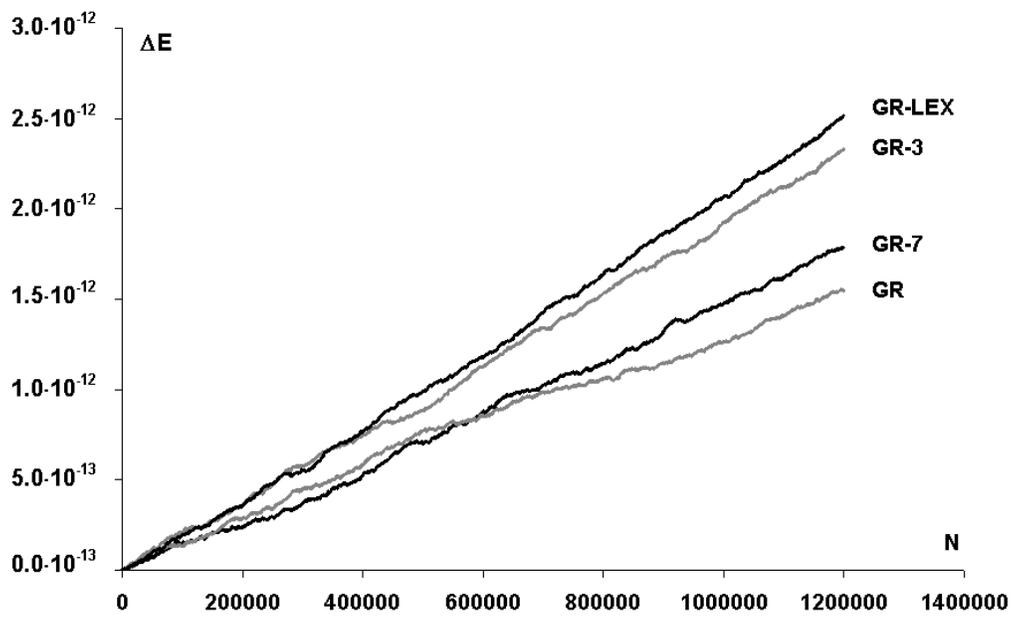}    
\caption{Energy error as a function of time ($t=h N$, $h=0.25$), for the simple pendulum, $p_0 = 1.8$ ($E_{ex} = 0.62$). }
\label{Egrad}      
\end{figure*}

Theoretically all gradient schemes \rf{graddel} preserve exactly the energy but, of course,  round-off errors cause some small inaccuracy, see Fig.~\ref{Egrad}. We see, that the energy error accumulates slowly, almost linearly but with a very very small slope: for $t \approx 300\ 000$ we have $\Delta E \approx 10^{-12}$.

\subsection{Stability and relative error of the period}

\begin{figure*}
  \includegraphics[width=\textwidth]{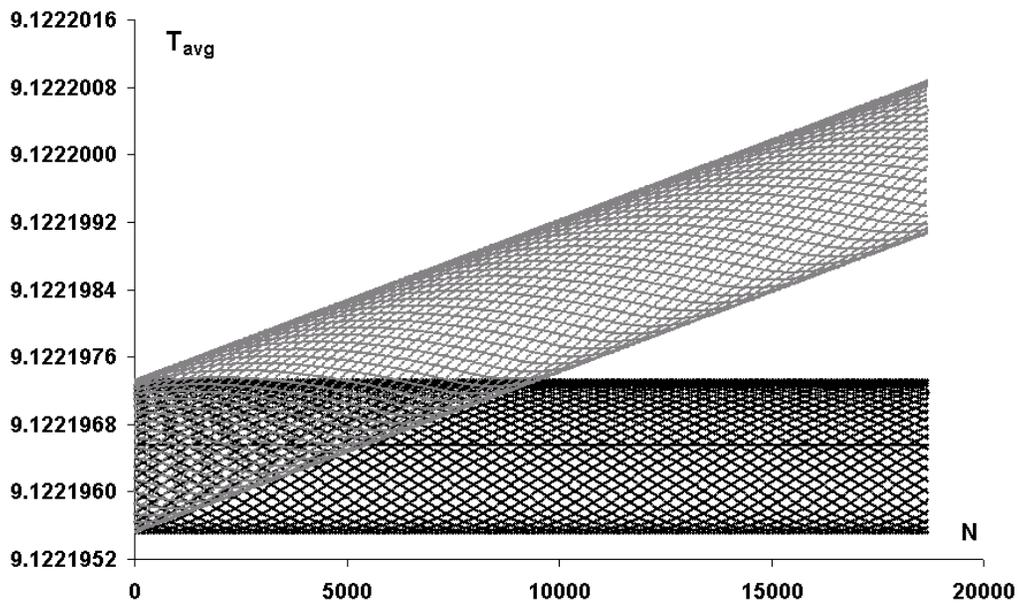}  
\caption{Average period as a function of time ($N$ is a number of half-periods) for the simple pendulum, $p_0 = 1.8$ ($T_{th} = 9.122\ 196\ 55$). Dark points -- GR-7, light points -- TAY-10, solid straight line -- exact period.  }
\label{stab-gr-tay}       
\end{figure*}

\begin{figure*}
\includegraphics[width=\textwidth]{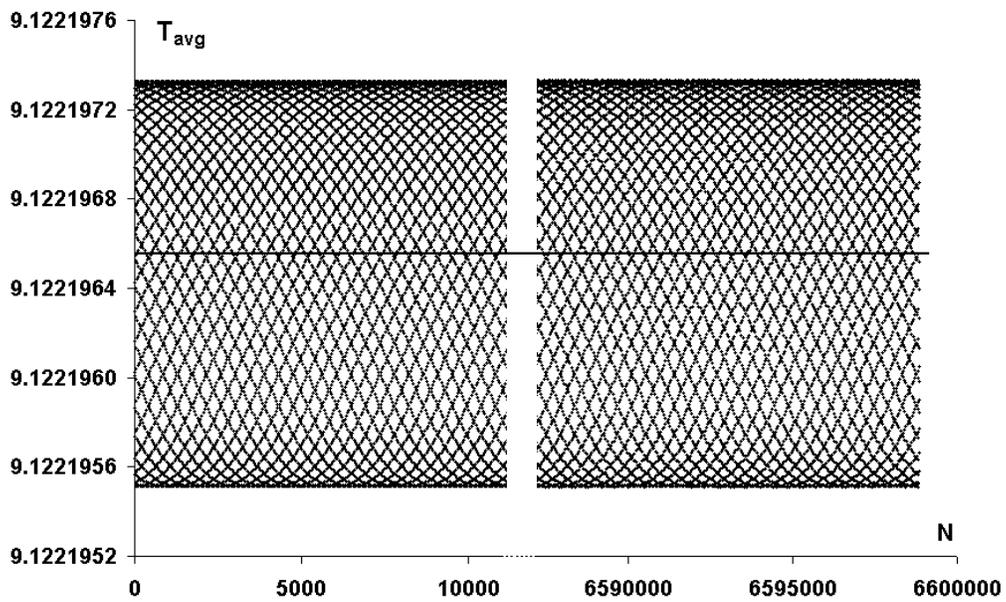}    
\caption{Average period as a function of time ($N$ is a number of half-periods) for the simple pendulum, $p_0 = 1.8$, scheme GR-11. Solid straight line -- exact period ($T_{th} = 9.122\ 196\ 55$).   }
\label{stab-gr11}       
\end{figure*}

All gradient schemes have extremaly stable period of oscillations. The stability of the discrete gradient scheme (GR) was tested in detail in \cite{CR-long}. Other gradient schemes follow the same pattern. Fig.~\ref{stab-gr-tay} compares the average period (more precisely: $T_{avg} (N,20)$,  see \cite{CR-long}) of numerical solutions produced by GR-7 and TAY-10. If $t$ is not very large, then in both cases the average period oscillates around the exact value $T_{th}$. For longer times we clearly see that TAY-10 becomes less and less exact, see Fig.~\ref{stab-gr-tay}, while   GR-7 oscillates exactly in the same way, even for very very long times, e.g., $t \approx 30\ 000\ 000$ at Fig.~\ref{stab-gr11}.

\begin{figure*}
  \includegraphics[width=\textwidth]{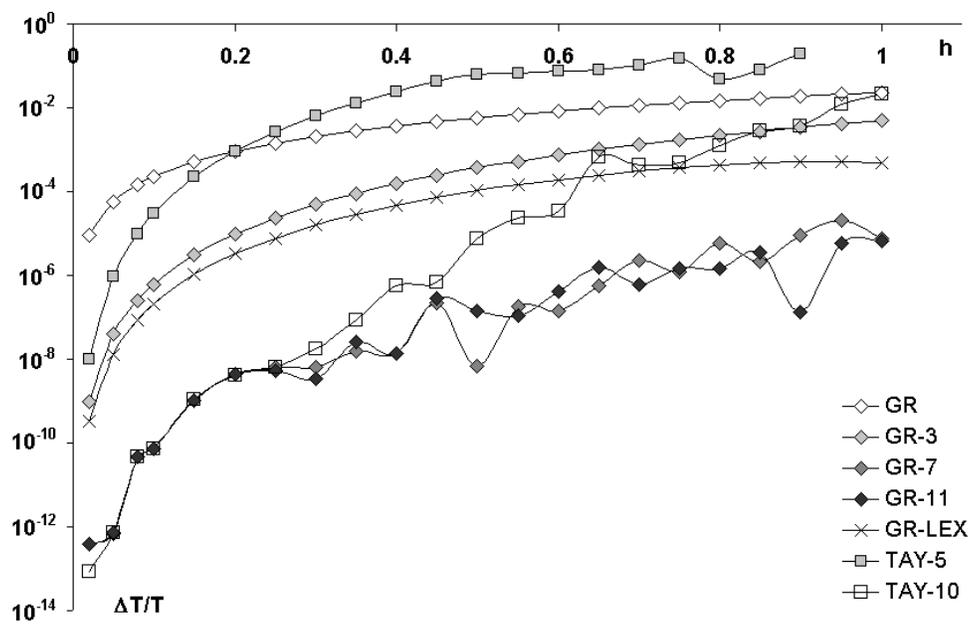}  
\caption{Relative error of the period of the simple pendulum as a function of $h$, for $p_0=1.95$ ($T_{th} = 11.657\ 585\ 28$).   }
\label{dT-195}       
\end{figure*}

\begin{figure*}
  \includegraphics[width=\textwidth]{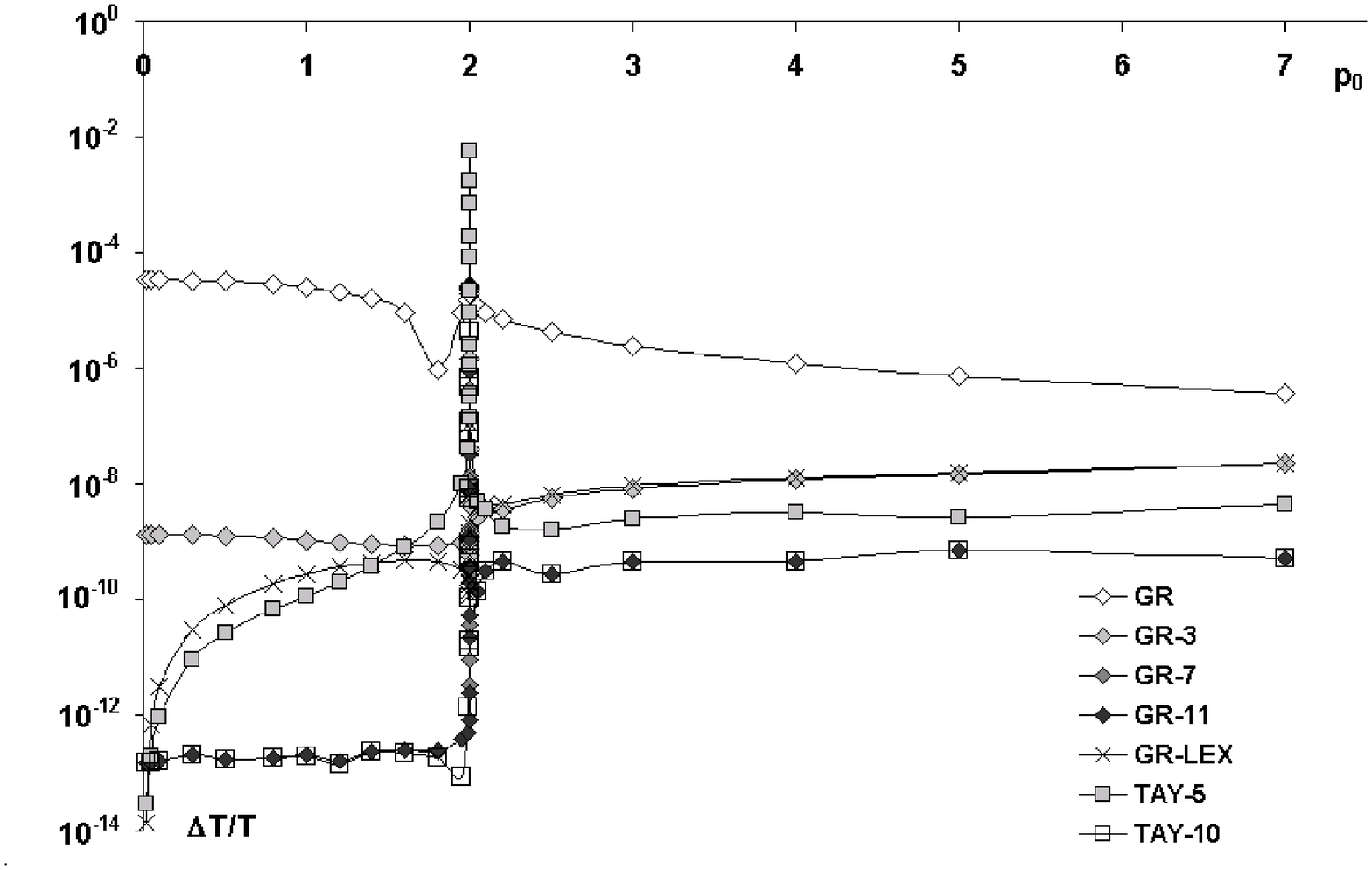}  
\caption{Relative error of the period for the simple pendulum as a function of $p_0$, for $h=0.02$.   }
\label{dT-h002}       
\end{figure*}

Fig.~\ref{dT-195} and Fig.~\ref{dT-h002} illustrate the relative error of the period. More precisely, we consider ${\bar T}_{avg} (0,100,200)$ (similarly as in \cite{CR-PRE}),   for details see \cite{CR-long}, p.\ 11 (roughly saying, we consider the first 200 periods making some averaging). Then, we compare the results with the exact period $T_{th}$. 

Fig.~\ref{dT-195} presents the dependence of the relative period on the time step $h$. We see that GR-7 yields excellent results  
(better by 3-4 orders of magintude than GR-LEX). The accuracy of TAY-10 and GR-11 is  (for $p_0 = 1.95$ and $h < 0.3$) almost the same. The accuracy of Taylor schemes becomes relatively lower for greater $h$. 
 
Increasing the order of GR-$N$ for small $h$ we increase the accuracy but only to some extent, see Fig.~\ref{dT-h002}.  Indeed, for $h=0.02$ schemes GR-11 and TAY-10 yield practicaly the same accuracy as GR-7, i.e., $10^{-13}$ for oscillating motions and $10^{-9}$ for rotating motions, with exception of the region $p_0 \approx 2$, where the accuracy is lower for any numerical scheme. For $p_0 < 2$ (oscillations) GR-7 is more accurate than GR by 7-9 orders of magnitude. For  small $p_0$ also  GR-LEX and TAY-5 attain such high accuracy. For $p_0 > 2$ the scheme GR-LEX produces almost the same results as GR-3, and both are less accurate than TAY-5 (GR-LEX is more and more accurate for decreasing $p_0$). We point out that gradient schemes produce very stable results (i.e., the picture presented at Fig.~\ref{dT-h002} is time-independent. The accuracy of Taylor schemes  decreases with time, see Fig.~\ref{stab-gr-tay}.

\subsection{Neighbourhood of the separatrix}

\begin{figure*}
\includegraphics[width=\textwidth]{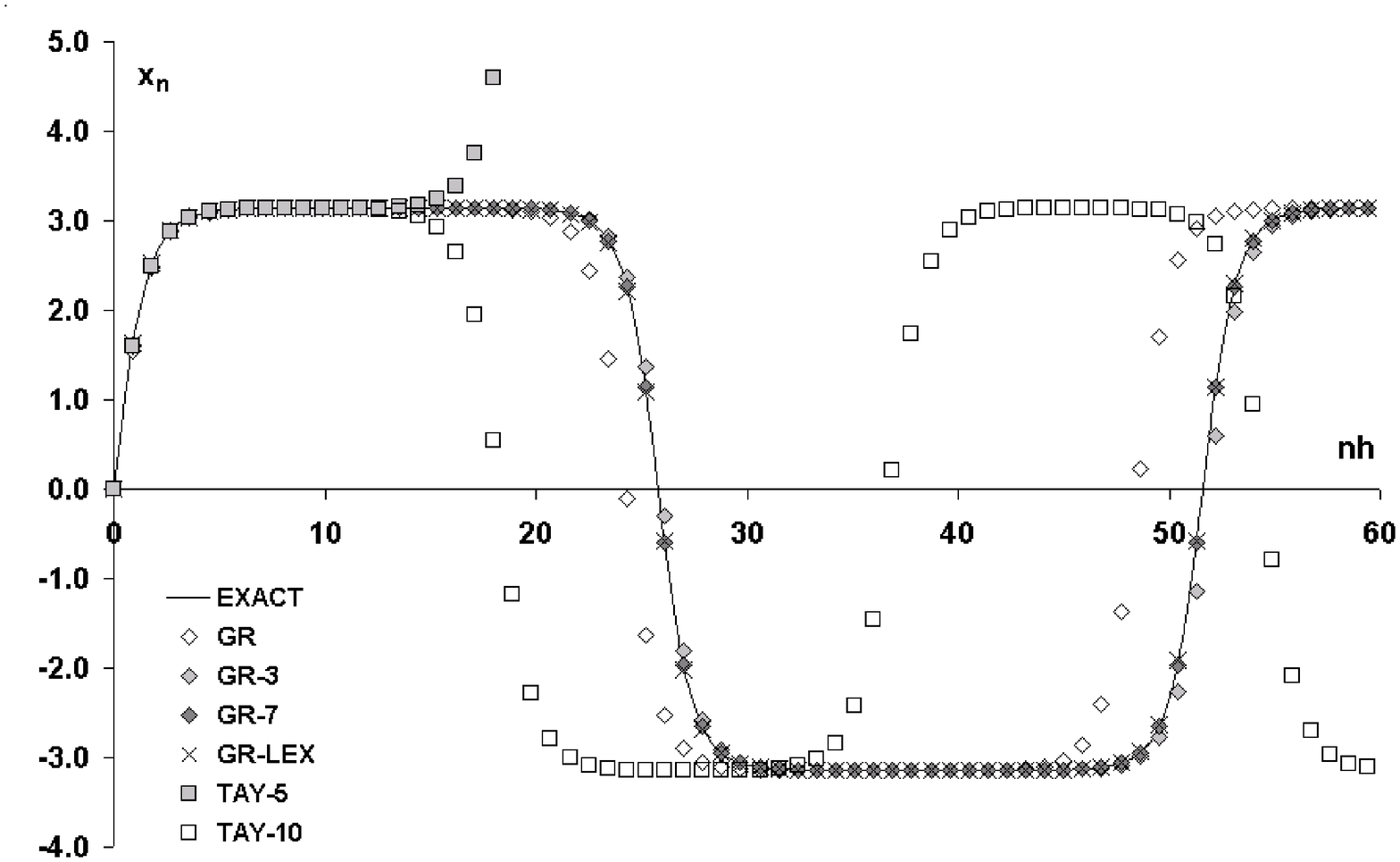}    
\caption{$x_n$ as a function of time ($t = n h$), very near the separatrix ($p_0 = 1.999\ 999\ 999\ 9$), $h=0.09$ for TAY-5, $h = 0.45$ for TAY-10, $h=0.9$ for all other discretizations. The solid line corresponds to the exact solution ($T_{th} = 51.596\ 879\ 14$).  }
\label{sep-2}       
\end{figure*}

The neighbourhood of the separatrix ($p_0 \approx 2$ for the simple pendulum) is most difficult to be simulated numerically. The discrete gradient method (GR) turns out to be relatively good in this region, see \cite{CR-long}, and the locally exact methods (GR-LEX, GR-SLEX) work almost perfectly \cite{CR-PRE}. 
Here, we take for comparison also GR-3, GR-7, GR-11 and TAY-5, TAY-10. The Taylor schemes are much worse in this region: for $h=0.9$ even TAY-10 is not able to reproduce the correct qualitative behaviour, see Fig.~\ref{sep-2}.   Throughout the first period  the scheme GR yields good qualitative behaviour and is better than TAY-10 with a halved time step, see Fig.~\ref{sep-2}.  In the first period  GR-3,  GR-7 and  GR-LEX (and also GR-SLEX and GR-$N$ for $N>3$) produce similar results.  We point out that the exact trajectory is very close to the separatrix ($|p_0 - 2| = 10^{-10}$) and $h$ is very large but, nevertheless, all improved discrete gradient methods simulate very accurately the motion of the pendulum. 

\begin{figure*}
\includegraphics[width=\textwidth]{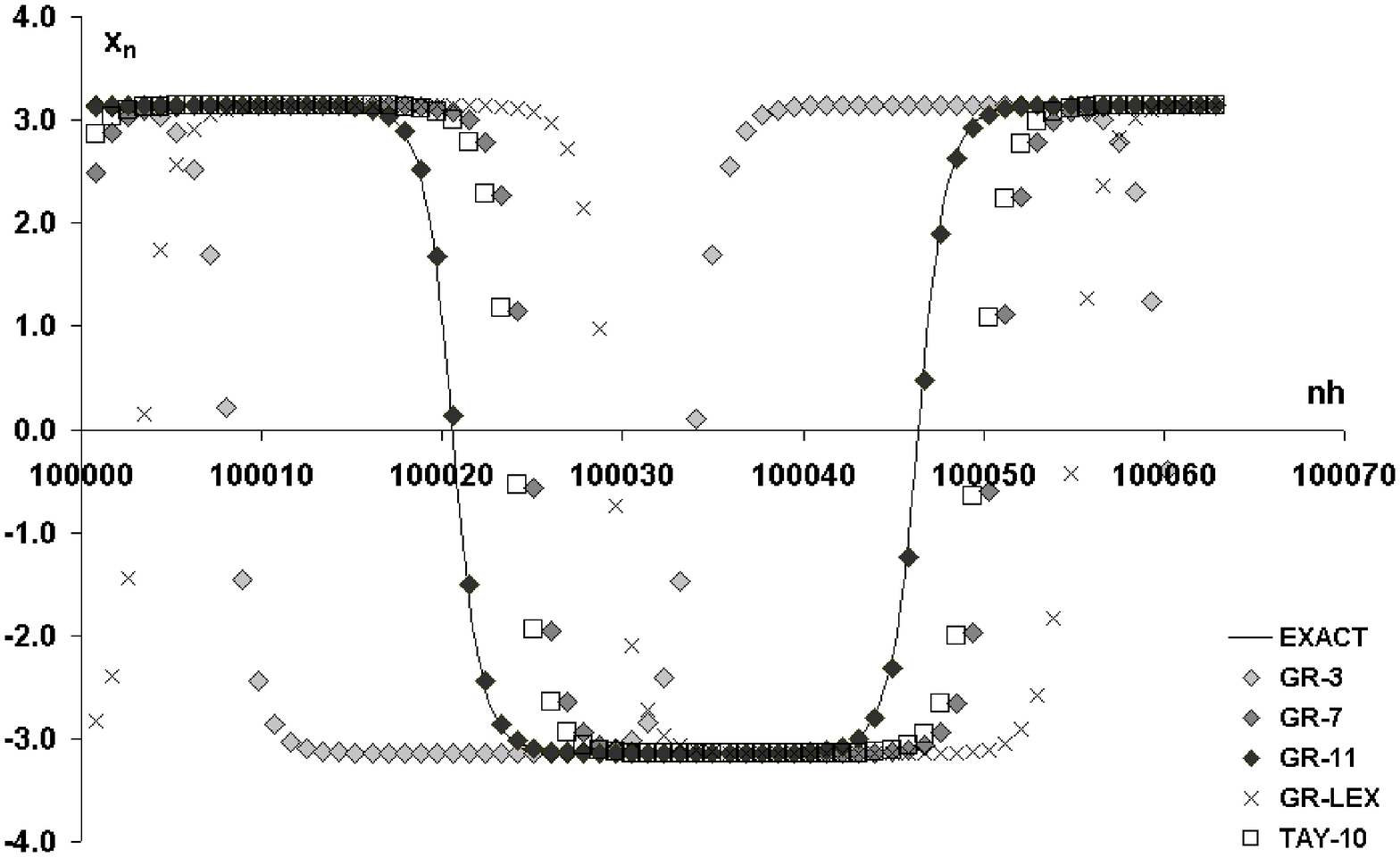}    
\caption{$x_n$ as a function of time ($t = n h$), very near the separatrix ($p_0 = 1.999\ 999\ 999\ 9$), $h=0.09$ for TAY-10 and $h=0.9$ for all other discretizations. The solid line corresponds to the exact solution ($T_{th} = 51.596\ 879\ 14$).  }
\label{sep-far}       
\end{figure*}

Fig.~\ref{sep-far} shows the same situation for much longer times ($t > 100\ 000$).  Note that the time step for TAY-10 is much smaller ($h=0.09$) than the time step for all gradient schemes, which is very large ($h=0.9$). In spite of that essential handicap, TAY-10 is only slightly better than GR-7 and less accurate than GR-11. GR-7 is more accurate than GR-LEX.

\section{Conclusions}

The numerical integrators GR-$N$, described in this paper,  have similar advantages as GR-LEX and GR-SLEX:  they preserve exactly the energy integral (i.e., eq.\ \rf{energy} holds), are extremaly stable and have very good long-time behaviour of numerical solutions. 
They can be constructed for any prescribed order $N$.

Therefore, modifications presented in this paper essentially improve the discrete gradient method (at least in the one-dimensional case) keeping all its advantages. Schemes GR-$N$ (for $N \geqslant 7$) are much more accurate than GR-LEX for most choices of parameters. Only in the region of small $p_0$ the scheme GR-LEX is comparable with discrete gradient methods of high order. 

We point out that numerical schemes \rf{graddel}, like all discrete gradient methods, are neither symplectic nor volume-preserving.  Moreover, schemes GR-$N$ are not time-reversible. 
Therefore, the conservation of the energy integral plus high order seem to be sufficient to assure oustanding qualitative and quantiaive properties of these methods.

\ods

\noindent {\it Acknowledgments.} 
 This research work has been supported by the grant No. N~N202 238637 from the Polish Ministry of Science and Higher Education.

\pagebreak

\end{document}